\documentclass[a4paper,11pt]{article}
\pdfoutput=1 % if your are submitting a pdflatex (i.e. if you have
\usepackage{jcappub} % for details on the use of the package, please
\usepackage[T1]{fontenc} % if needed
\usepackage{footnote}

\title{\boldmath{Cosmological birefringence constraints from CMB and astrophysical polarization data}}

\author[a]{M. Galaverni,}
\author[b,c]{G. Gubitosi,}
\author[d]{F. Paci}
\author[e,f]{and F. Finelli}
\affiliation[a]{Studio Teologico Interdiocesano,\\
V.le Timavo 93, 42121 Reggio Emilia, Italy}
\affiliation[b]{Dipartimento di Fisica and sez. Roma1 INFN, Universit\`a di Roma ``La Sapienza'',\\P.le A. Moro 2, 00185 Rome, Italy}
\affiliation[c]{Theoretical Physics, Blackett Laboratory, Imperial College, London, SW7 2BZ, U.K.}
\affiliation[d]{SISSA, Scuola Internazionale Superiore di Studi Avanzati,\\Via Bonomea 265, Trieste, 34136, Italy}
\affiliation[e]{INAF-IASF Bologna, via Gobetti 101, I-40129 Bologna, Italy}
\affiliation[f]{INFN, Sezione di Bologna, Via Irnerio 46, I-40126 Bologna, Italy}

\emailAdd{matteo.galaverni@gmail.com}
\emailAdd{giulia.gubitosi@imperial.ac.uk}
\emailAdd{fpaci@sissa.it}
\emailAdd{finelli@iasfbo.inaf.it}

\abstract{
Cosmological birefringence is a rotation of the polarization plane of photons coming from sources 
of astrophysical and cosmological origin.  
The rotation can also depend on the energy of the 
photons and not only on
the distance of the source and on the cosmological evolution of the
underlying theoretical model.
In this work, we constrain few selected models for cosmological birefringence, 
combining CMB and astrophysical data at radio, optical, X and $\gamma$ wavelengths, taking into account  
the specific energy and distance dependences. 
}

\begin{document}
\maketitle
\flushbottom

\section{Introduction}
\label{sec:intro}

The precision in the measurements of light polarization has significantly improved 
over a wide range of frequencies (from few GHz to $10^9$ GHz, from few $\mu$eV to hundreds of keV) in the last couple of decades.
From the first observations at radio and optical wavelengths, 
there has been a tremendous advance in the measurement of polarization from distant sources, passing through the 
detection of the polarization of the Cosmic Microwave Background (CMB) \cite{Kovac:2002fg}.  
Measurement of light polarization from  distant sources can therefore be a powerful probe of theories beyond the Standard Model predicting modifications 
in the photon dispersion relation.

From a theoretical perspective, several extensions of the standard Maxwell electromagnetic theory 
predict a rotation in the polarization plane of light, which 
accumulates over cosmological distances to become potentially observable (cosmological birefringence).
Carroll, Field and Jackiw studied how the addition of a Chern-Simons term $\propto p_\mu A_\nu \tilde{F}^{\mu \nu}$ 
produces a wavelength independent rotation of linear polarization which could be constrained by astrophysical data \cite{Carroll:1989vb}. 
The signature in the power spectra of CMB polarization anisotropies is independent of the angular multipole for a timelike vector 
$p_\mu$ \cite{Lue:1998mq}. 
The coupling to pseudo-scalar fields $\propto \phi F_{\mu \nu} \tilde{F}^{\mu \nu}$ could also imply 
a frequency independent rotation of the polarization plane generated by the coupling to photon 
\cite{Carroll:1991zs,Harari:1992ea,Carroll:1998zi, Finelli:2008jv}. The corresponding effect on the CMB polarization power spectra now depends on the 
multipole in general \cite{Liu:2006uh,Finelli:2008jv}.
Other theories, as Weyl-type Lorentz breaking models, predict a modification in the orientation 
of polarization increasing linearly with energy \cite{Shore:2004sh}. 
A cosmological birefringence effect proportional to the square of the photon energy  
emerges in the effective field theory approach to Lorentz violation \cite{Myers:2003fd}.

In the early nineties, radio galaxies were among the few astrophysical sources for which polarization was measured. 
For this reason, the first tests of birefringence were based on the radio \cite{Carroll:1989vb} and the optical bands \cite{Cimatti:1993yc}.
In those years a claim for the detection of an energy independent cosmological birefringence\cite{Nodland:1997cc} 
did not pass further verifications \cite{Carroll:1997tc,Leahy:1997wj}. 
Today, it is possible to obtain interesting constraints also considering other types of sources.
In particular, the available data on the CMB anisotropies polarization 
is now significantly increased.
This provides a very good test for cosmological birefringence since a rotation of the linear
polarization plane modifies the polarization pattern mixing B modes and E modes in a peculiar way. 
There is indeed a very large amount of works using CMB to constrain birefringence (see \emph{e.g.} 
\cite{Feng:2006dp, Cabella:2007br, Xia:2007qs, Finelli:2008jv, Gubitosi:2009eu, Pagano:2009kj, Gruppuso:2011ci, Barkats:2013jfa, Gubitosi:2014cua}).
At higher energies the UV emission from a sample of radio galaxies has been considered as a probe of a cosmological rotation of the 
polarization plane \cite{Alighieri:2010eu},
also combined with CMB observations \cite{Alighieri:2010pu,Kaufman:2014rpa}\footnote{Note that these works use rotation angles from different data without
considering the different conventions by the CMB and astronomical
community, an important point of this work.}.
At much higher energies (hundreds of keV)
hard X-ray band observations of the Crab Nebula polarization \cite{Dean:2008zz}
provide an important test for birefringence \cite{Maccione:2008tq}. 
%At such high energies, also 
Polarization measurements for $\gamma$-ray bursts (GRBs) are also used in this context \cite{Fan:2007zb,Stecker:2011ps, Toma:2012xa,Gotz:2013dwa,Laurent:2011he}.

Given the variety of different observations that have been used for constraining birefringence, 
it is interesting to study what conclusions can be drawn from a combined analysis of different data. 
This would in principle give  more reliable information on  the effect under scrutiny.  However, combining the constraints is not a trivial task: as we mentioned, different models for birefringence predict different energy dependence of the effect, and also different dependence on the photon travel time. So necessarily the analysis has to be model-dependent, in order to combine data so that energy and distance of each source are properly accounted for. 
The energy-dependence issue was already pointed out in \cite{Gubitosi:2012rg}, where different CMB datasets were combined in order to test the energy dependence of birefringence. When combining observations of different sources, the additional difficulty arises of taking into account the distance dependence as well. This issue is addressed in this work.

Note that some models predict a non-isotropic effect \cite{Carroll:1989vb, Gubitosi:2010dj}. Tests of anisotropic birefringence are more complicated: for point-like sources, one needs a large set of objects with a statistically significant spatial distribution \cite{Kamionkowski:2010ss}. CMB data would seem in principle more competitive in this respect due to their almost complete coverage of the sky, and some efforts have been made already in this direction \cite{Gruppuso:2011ci,Yadav:2009eb, Gubitosi:2011ue, Gluscevic:2012me}. %However, the present instrumental sensitivity does not allow to put very strong constraints. 
In this work we will constrain  birefringence effects assuming isotropy.

Our paper is organized as follows.  
Section II provides a review of the astronomical and cosmological
conventions for the measurement of polarization position angles.
We also present the different datasets used: CMB, UV distant radio
galaxies, radio sources,
Crab Nebula and $\gamma$-ray bursts.
In Sec. III, for each model considered, we combine these datasets - extending
the work by two of us for CMB only \cite{Gubitosi:2012rg} -
by taking into account the peculiar energy and distance dependence. We estimate our constraints and  
%We discuss the constraining power of each of the datasets depending on the model considered, showing that which dataset dominates the results is strongly model-dependent.
discuss the relative constraining power among our datasets for each theoretical model.
Finally, we summarize and conclude in Sec. IV.

In this work, we use natural units, $\hbar=c= 1$, and assume a cosmological model with 
{\em Planck} 2013 estimates of cosmological parameters \cite{planck2013-p11} $H_0=h \cdot 100$ Km/s/Mpc = 67.2 Km/s/Mpc, 
$\Omega_\mathrm{M} = 0.32$ and $\Omega_\Lambda = 0.68$\,. The amount of rotation of the polarization plane is denoted by $\alpha$.

\section{Dataset}
\label{sec:data}

In this section we present the datasets currently used to 
constrain cosmological birefringence using the observation of polarization coming from different astrophysical and cosmological sources. 
Since different astronomical communities are involved in this kind of measurements, there is no common convention for the sign of the rotation angle of the polarization plane. So, we first recall what conventions have been used by the different communities and we fix a common one to be applied when combining  the datasets.

\subsection{Polarization convention}
\label{sec:conv}

The cosmological \cite{Hamaker2004} and astronomical communities 
(IAU convention \cite{HamakerBregman1996}) have adopted opposite conventions for measuring the  polarization plane direction
(see also \cite{Ade:2014gna}).
The CMB comunity convention follows from associating a Cartesian reference frame to each point of the sky such that the $x$ and $y$ axes point towards the {\it South} and {\it East} respectively, and the $z$ axis points away from the observer (ie, outwards). The polarization angle increases clockwise looking at the source. 
Instead, the reference frame for the astrophysical community is defined by the $x$ and $y$ axes pointing respectively toward the {\it North} and {\it East}, and the $z$ axis pointing toward the observer (ie, inwards). The polarization angle increases anti-clockwise looking at the source.
The different convention used reflects onto a change of sign in the $U$ Stokes parameter and in the polarization angle.

In this work, we adopt the CMB convention. Therefore, the constraints on $\alpha$ reported in the literature from 
CMB observations will be taken as they are, whereas we change sign to the values of $\alpha$ obtained in the astrophysical conventions. 

\subsection{CMB}

CMB data provide constraints on birefringence associated to radiation with low energy but arriving from the 
largest distances, so they can become important in constraining models where the amount of rotation is an 
increasing function of propagation time and does not increase much with energy.

Rotation of CMB linear polarization between the last scattering surface ($z_{\mathrm {decoupling}}=1090.43\pm 0.54$ at $68 \%$ C.L. as from \citep{planck2013-p11}) 
and the observer 
modifies the gradient and curl of the
polarization pattern ($E$ and $B$ modes following \cite{Zaldarriaga:1996xe}), mixing the two modes in a characteristic way.
In Table~\ref{tab:CMB} we report the most up-to-date constraints on the
birefringence angle $\alpha$ from CMB polarization experiments as found in  \cite{Gubitosi:2012rg}, updated with the last WMAP9 results \cite{Hinshaw:2012aka}.
Since we are interested in studying the energy dependence of the effect, we also note down the characteristic effective energy measured by each experiment.

\begin{table*}[htbp]
\caption{CMB constraints for cosmological birefringence (68\% C.L.).
For BOOMERANG the error already takes into account the
systematic error of $-0.9\pm0.7$ deg.
}
\begin{center}
\begin{tabular}{|c|c|c|c|}
\hline
Experiment      & Energy (GHz)   &  $\alpha\pm stat (\pm syst)$ (deg) & Reference  \\
\hline
\hline
WMAP9 & 53 & $-0.36\pm 1.24 (\pm 1.5)$ & \cite{Hinshaw:2012aka}   \\
BOOM03 & 145 & $-4.3\pm4.1$
 &  \cite{Pagano:2009kj} \\
BICEP1 & 129 & $-2.77\pm0.86 (\pm 1.3)$ & \cite{Kaufman:2013vbd}  \\
QUAD & 100 & $-1.89\pm2.24 (\pm 0.5)$ & \cite{Wu:2008qb}  \\
QUAD & 150 & $0.83\pm0.94(\pm 0.5)$ & \cite{Wu:2008qb}  \\
\hline
\end{tabular}
\end{center}
\label{tab:CMB}
\end{table*}
For our purposes, it is enough to use the constraint from WMAP9.
By adding  in quadrature statistical and systematic errors  for WMAP9, we obtain a constraint on cosmological birefringence  $\alpha_{\rm CMB}=-0.36\pm 1.9$ deg for photons observed at an energy of 
$2.2\times 10^{-4}$ eV.

\subsection{UV emission from radio galaxies}

At much lower redshifts ($2\lesssim z \lesssim 4$), predicted perpendicularity between the 
UV linear polarization and the direction of the UV axis in radio
galaxies provides another constraint on cosmological birefringence \cite{Cimatti:1993yc}.
Due to energy redshift, distant radio galaxies photons emitted in UV 
are observed in optical wavelengths, $\lambda_0\sim 500\,\mbox{nm}$, $E_0\sim 2.5$ eV. 
In Table~\ref{tab:Pol_gal_tab}, we report the most recent constraints
on cosmic birefringence, derived in  
\cite{Alighieri:2010eu}. We also provide their average distance and the rotation angle. Both averages are weighted with the errors on $\alpha$.

\begin{table*}[!h]
\caption{Constraints on the rotation of the linear polarization plane for distant UV radio galaxies (RG) \cite{Alighieri:2010eu}.}
\begin{center}
\begin{tabular}{|c|c|c|}
\hline
RG name      &  $z$     & $\alpha \pm \Delta\alpha \;\rm{[deg]}$\\
\hline
\hline
MRC 0211-122 & 2.34 &  $1\pm 3.5$\\
4C -00.54    & 2.363 & $8\pm 8$\\
4C 23.56a    & 2.482 &  $-4.6\pm 9.7$\\
TXS 0828+193 & 2.572  & $-1.6\pm 4.5$\\
MRC 2025-218  & 2.63   & $4\pm 9$\\
TXS 0943-242  & 2.923  & $0.3\pm 4.4$\\
TXS 0119+130  & 3.516 & $-5\pm 16$\\
TXS 1243+036  & 3.570  & $4.0\pm8.8$\\
\hline
\hline
Mean & 2.62  & $0.7\pm 2.1$ \\
\hline
\end{tabular}
\end{center}
\label{tab:Pol_gal_tab}
\end{table*}

\subsection{Radio sources}

Another constraint on  cosmological birefringence can be derived
from the relation between polarization and total intensity structures of radio galaxies and quasars \cite{Leahy:1997wj}.
This provides the constraint at lowest energy, from relatively close-by sources.

In Table~{\ref{tab:radio}}, we reproduce the constraints
on the birefringence angle $\alpha$ found in \cite{Leahy:1997wj} and obtained with images at 
$\lambda=3.6$ cm ($E_0\simeq 3.4\times 10^{-5}$ eV). 
Also here we  provide the average source distance and the average rotation angle, where both averages have been weighted with the uncertainty on $\alpha$.
We have removed one source from the original list, as its redshift is significantly higher than the others. 
This is to ensure a set of sources characterized by a relatively homogeneous distribution of distance, being the distance itself accounted for in our analysis. 

\begin{table*}[!htbp]
\caption{Constraints on the birefringence angle $\alpha$
from selected radio sources \cite{Leahy:1997wj}.}
\begin{center}
\begin{tabular}{|c|c|c|}
\hline
Name      & $z$   &  $\alpha\pm \Delta\alpha$ [deg]  \\
\hline
\hline
3C 34 & 0.6897 & $-8\pm12$  \\
3C 47 S & 0.425 & $2\pm2$ \\
3C 55 & 0.735 & $-6\pm9$ \\
3C 228 & 0.5524 & $13\pm20$ \\
3C 244.1 & 0.428 & $1\pm9$ \\ 
3C 265 & 0.8108 & $0\pm10$  \\ 
3C 268.1 & 0.97 & $-13\pm20$ \\ 
3C 330 & 0.428 & $5\pm12$  \\ 
3C 340 & 0.7754 & $6\pm9$ \\
\hline
\hline
Mean & 0.47& $1.6\pm 1.8$ \\
\hline
\end{tabular}
\end{center}
\label{tab:radio}
\end{table*}

\subsection{Crab Nebula}

At much higher energies than the ones considered until now, constraints based on the observation of the Crab Nebula have been set, for which we refer to \cite{Maccione:2008tq}.  The result is based on the comparison between the neutron star rotation axis - measured by HST 
and Chandra satellites - and the gamma-ray polarization direction - observed by INTEGRAL -, expected to be aligned.
The former axis lies at a position angle $\varphi=124.0^\circ\pm0.1^\circ$, whereas the latter at $\varphi=123^\circ\pm11^\circ$, 
measured from North anticlockwise
within a energy band from 150 to 300 keV. 
Therefore, we assume the difference of the two as an estimate of the birefringence angle, $\alpha=1\pm11$ in the CMB convention. The distance of the Crab Nebula is 1.9Kpc, corresponding to $z=4.5\times 10^{-7}$. So this is the 
closest source we include in our analysis and also the one at the highest frequency.

\subsection{Gamma-ray bursts}

Constraints 
from GRBs necessitate a separate analysis.
Here we can not rely on a direct measurement of the cosmological
birefringence, but limits are derived from linear polarization
measurements at different energies. 
Therefore we can not use these data directly to infer 
the energy dependence of cosmological birefringence, 
but once we have selected a particular model 
(except the energy independent case) 
they provide quite stringent constraints (see Sec. III).

Limits are derived in two different ways. 
Since detectors have a finite bandwidth ($k_1<E<k_2$), an order of magnitude of the effect can be obtained by looking 
at the total degree of polarization: if the rotation of linear polarization angle were larger than
$\pi/2$ over the detector energy range, a polarization loss would be produced \cite{Gleiser:2001rm}.
The other method relies on the  comparison of the polarization direction at different energies.
%Both methods assume  the effect to be energy-dependent, both being based on the comparison of different energy channels. 
In this work, we will only rely on the second one
 following \cite{Gotz:2013dwa}.
%since it provides a more direct test of birefringence.

Direct polarization measurements at such high energies are very difficult.  
Several claims have been made in the past \cite{Coburn:2003pi},
although results were refuted by an independent data analysis \cite{Rutledge:2003wa}.
So, in general one should be cautious before drawing strong conclusions based on the GRB data only. 
However, the number of observations is continuously increasing
\cite{Toma:2012xa,Laurent:2011he}, and the techniques are improving, so that it is still worth exploring the potentialities of this method. Here, in order to give an idea of this, we will provide results both including and not including the GRB data.

We refer to \cite{Gotz:2013dwa} as an example of the GRB capabilities, as this provides the latest results.
The polarization direction for GRB061122 is measured in two different energy bands, $250-350$ keV and $350-800$ keV, obtaining, respectively, $\phi_{1}=145\pm 15$ and $\phi_{2}=160\pm 20$ at $68\%$ C.L..
The distance of the GRB source is given as $z=0.54$.
As done in \cite{Gotz:2013dwa}, we will use the conservative constraint on the rotation angle $\alpha=0\pm 50$ degrees (68\% C.L.). 
Following \cite{Gubitosi:2012rg} we introduce an 
effective energy depending on the functional dependence of $\alpha$ on energy and on the bandwidth of the different channels:
$E= 530$ keV for the linear dependence 
and $E=550$ keV in the quadratic case.

\section{Analysis and results}
\label{sec:res}

In this Section we constrain different models predicting birefringence, each one characterized by a different energy and distance dependence. %We combine the datasets previously introduced (see Table~\ref{tab:total} for a summary), extending the work of \cite{Gubitosi:2012rg}.

\begin{figure}[!htdp]
\includegraphics[scale=0.59]{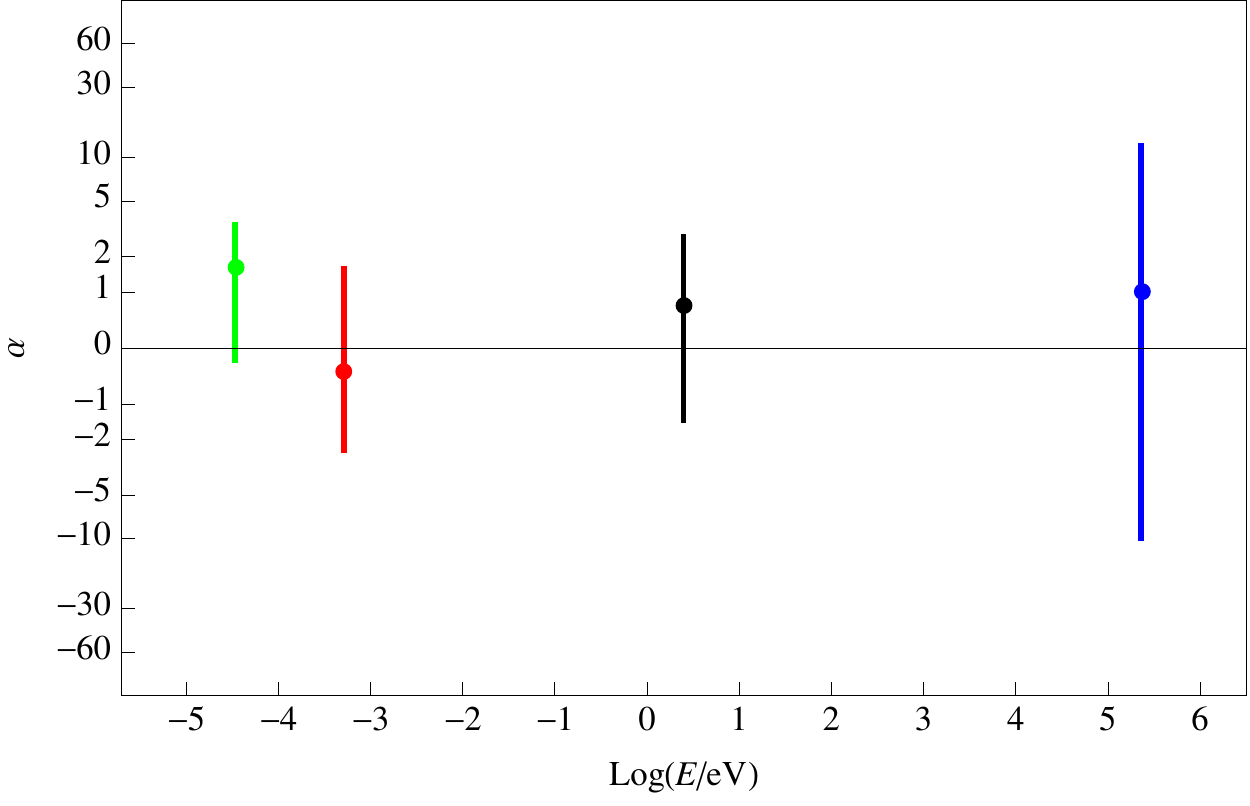}
\includegraphics[scale=0.59]{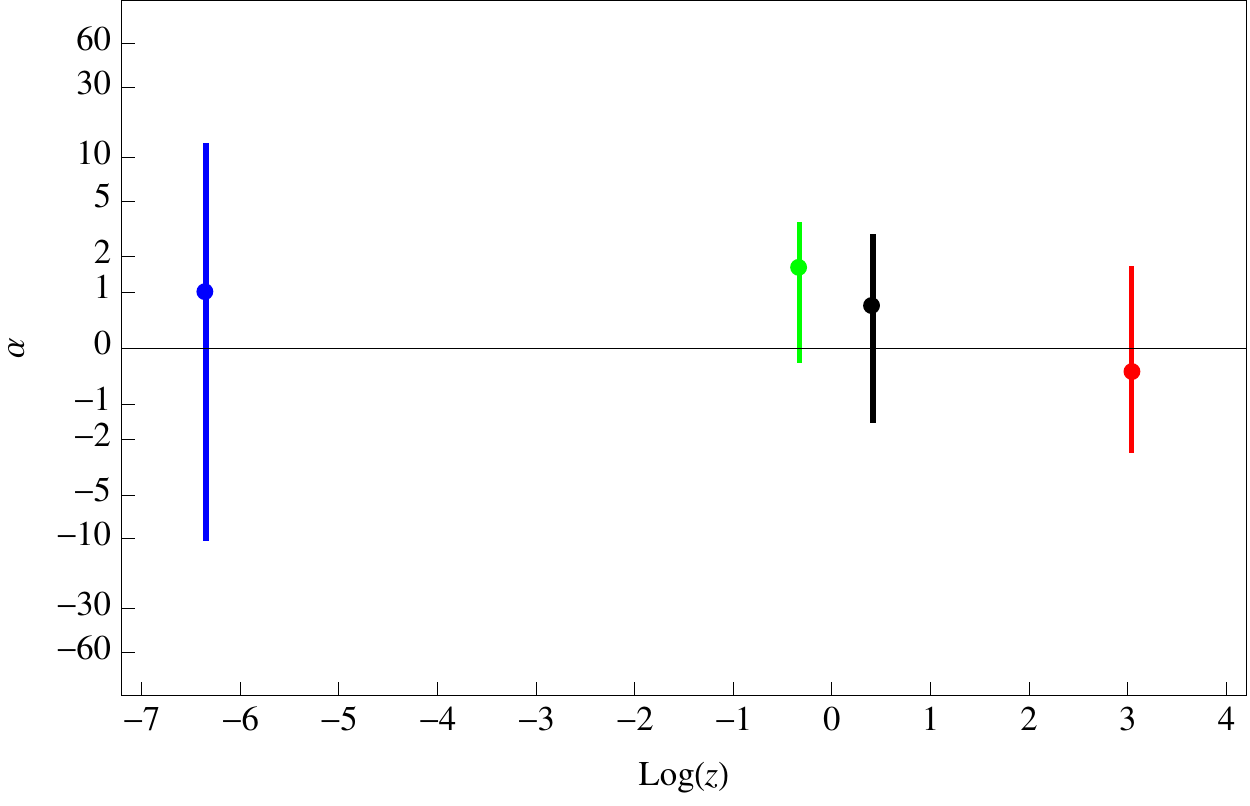}
\caption{Energy (left panel) and redshift (right panel) dependence of constraints for the cosmological birefringence angle $\alpha$;
black, green, red and blue points refer to UV radio galaxies, radio sources, CMB, Crab Nebula. Note that the $\alpha$ axis is rescaled with a $\sinh$ function. This rescaling has the side effect of making error bars looking asymmetric, while they actually are not. Note that GRBs can be used to constrain birefringence only after assuming a particular energy dependence for $\alpha$ (linear or quadratic). For this reason we don't include the datapoint in the plots.}
\label{fig:datapoints}
\end{figure}

One of the main goal of our analysis is to combine birefringence constraints
over a very large energy range, extending the work of \cite{Gubitosi:2012rg}. 
In order to clarify the specific contribution of each dataset 
we decided to consider only a representative constraint 
for each single data set (see Table~\ref{tab:total} for a summary).
Whereas in \cite{Gubitosi:2012rg}  data were uniform in their distance distribution (they all referred to CMB observations), the sources considered in this work lie at very different distances from us. Accordingly, we take into account the expected propagation time dependence for each model we investigate. 
Fig. \ref{fig:datapoints} shows on the left panel the very wide frequency range explored, and on the right panel the redshift distribution of the sources.
Note that for CMB we use the reference redshift value $z_{\rm CMB}=1090$, compatible with $z_{\mathrm{decoupling}}$ introduced in the previous section \footnote{Note that including the uncertainties in energy and/or redshifts would be rather complicate since these errors are not provided in the literature in a consistent way. We also think that the dependence on energy and/or redshift of our main results in Fig. 1 are sufficiently smooth to ensure that additional uncertainties in energy and/or redshift could not alter significantly our conclusions.
}.
%Both in the plots and in the analysis we disregard error bars on energy and redshift of the sources, as they contribute  to the results in a negligible way 
%with respect to the uncertainties in $\alpha$. This is because we see that, if any dependence on energy and redshift is present, it is very weak 
%(see figure \ref{fig:datapoints}). So the errors on the horizontal axis of the plots would be suppressed when propagated to the vertical axis. 
As mentioned in the previous section, we present both the results obtained including and not including the GRB data point in our analysis. 
%In fact, although this turns out to provide the most stringent constraint in most cases, its reliability is still debated. 

\begin{table*}[!htbp]
\caption{Summary of the current constraints on the cosmological birefringence angle $\alpha$ coming from a variety of astrophysical and cosmological observations; for each dataset we report the typical redshift 
and the effective energy.
Note that GRBs can be used to constrain birefringence only after assuming a particular energy dependence for $\alpha$ (linear or quadratic). For this reason we don't include the datapoint in the plots.}
\begin{center}
\begin{tabular}{|l|c|c|c|}
\hline
Dataset     & $z$ & $E$ [eV]   &  $\alpha\pm\Delta\alpha$ [deg]  \\
\hline
\hline
CMB & 1090 & $2.2\times 10^{-4}$ & $-0.36\pm 1.9$  \\
UV RG & 2.62 & 2.5 & $0.7\pm2.1$  \\
Radio sources & 0.47 & $3.4\times 10^{-5}$ & $1.6\pm1.8$  \\
Crab nebula & $4.5\times 10^{-7}$  & $2.3\times10^5$ & 1$\pm$11   \\
\hline
Gamma-ray bursts (lin) & 0.54 & $5.3\times 10^5$ & 0$\pm$50   \\
Gamma-ray bursts (quad) & 0.54 & $5.5\times 10^5$ & 0$\pm$50   \\
\hline
\end{tabular}
\end{center}
\label{tab:total}
\end{table*}

Several authors have showed how the presence of a Chern-Simons term causes an
\textit{energy-independent} rotation of the polarization
plane \cite{Carroll:1989vb}. In this case the 
%(or cosmic polarization rotation following the recommendation  of \cite{Ni:2009fg}).
 amount of rotation is linearly dependent on the distance traveled by photons. 
 Another physical mechanism producing energy-independent birefringence is  a coupling between the electromagnetic field and a scalar (quintessential) field (see \emph{e.g.} \cite{Carroll:1998zi}). The amount of rotation would depend on the variation of the scalar field between emission and detection.  
These two theoretical scenarios have to be treated differently when combining sources located at significantly different distances. In particular, the latter would require to account for the cosmological evolution of the field if the distance of the sources plays a role, which is the case here. This kind of analysis goes beyond the scopes of the present paper, and is left to future work. Here, we will concentrate on the former case, an energy-independent birefringence effect linearly dependent on the propagation distance.
The expected amount of rotation can be written as \cite{Carroll:1989vb}:
\begin{equation}
\alpha=-\frac{1}{2}\,p_0 \,\Delta\ell
\end{equation}
where $\Delta\ell$ is the source distance and $p_0$ is the time-component of a fixed time-like vector which is coupled to the electromagnetic field \footnote{In principle, as done in \cite{Carroll:1989vb}, one could consider a general vector $p_\alpha$, but this would produce non-isotropic effects. As mentioned in the introduction, here we work under the assumption that birefringence is isotropic, so that only the time part of the vector could be present in the model.}. Taking into account the universe expansion, the rotation angle $\alpha$ can be written as a function of the source redshift $z_{\star}$ as:
\begin{equation}
\alpha(z_\star)=-\frac{1}{2}\, p_0 \int_{0}^{z_{\star}}\frac{1}{(1+z)H(z)}dz.
\end{equation}
Combining all data excluding the GRB data point, as this is obtained assuming that the effect is energy-dependent, we obtain:
\begin{equation}
p_0=(-0.93\pm2.9)\cdot 10^{-35} h \;\text{eV} \,,
\end{equation}
at 68\% C.L.. % where $h_0$ is the reduced Hubble constant.
Note that the dominant contribution to the result comes from the CMB and UV radio galaxies, 
because of their significantly higher distance than the others. 
We obtain indeed a constraint comparable to what found in \cite{Gubitosi:2012rg}, $|p_0|<9.4 \cdot 10^{-35} h \;\text{eV}\, \mathrm{at}\, 68 \%\, \mathrm{C.L.}$, where CMB data only were used.

\textit{Linear energy dependence} ($\alpha(E)\propto E$) can be due to the `Weyl' interaction described in \cite{Shore:2004sh} and \cite{Kahniashvili:2008va}.
In this case, neglecting energy redshift, the polarization rotation angle depends linearly on the distance traveled by photons, $\Delta \ell$,  and on the dimensionless scalar $\Psi_{0}$, which sets the amplitude of the interaction:
\begin{equation}
\alpha(E)\propto E \,\Psi_{0} \Delta\ell.
\end{equation}
When redshift effects are included
%, the dependence on the distance becomes non trivial, and it is easier to consider  the redshift dependence instead. In this case 
one has \cite{Gubitosi:2012rg}:
\begin{equation}
\alpha(E_0,z_\star)=8\pi E_0 \Psi_{0} \int_{0}^{z_{\star}}H(z)^{-1}dz\,,
\end{equation}
where $E_{0}$ is the photon energy today. This is the formula we use for our analysis.
Combining  the  first four datasets of Table~\ref{tab:total}, the best-fit value for $\Psi_{0}$ is:
\begin{equation}
\Psi_{0}=(3.0\pm 9.1)\cdot 10^{-37} h\,,
\end{equation}
at 68\% C.L.;  the dominant contribution comes from 
UV radio galaxies.
This is an interesting example of a case in which the dominant contribution does not come from the highest energetic source (the Crab Nebula), which is what one might naively expect. In fact, the distance dependence plays an important role and can compensate for the lower energy of other more distant sources. 
 If we include also the constraint from GRB,
this dominates and we obtain, at 68\% C.L.:
\begin{equation}
\Psi_{0}=(0.0\pm3.0)\cdot 10^{-40} h\,.
\end{equation}
In both cases the constraint improves by several orders of magnitude the estimate based on CMB data only, $\left|\Psi_{0}\right|<5.8\cdot 10^{-33} h$
\cite{Gubitosi:2012rg}.  

The \textit{quadratic energy dependence} ($\alpha(E)\propto E^2$) of the birefringence angle might be traced back 
to Quantum Gravity Planck-scale effects \cite{Myers:2003fd, Gubitosi:2009eu},  whose relics at low energies can be modeled as a coupling of the EM field with a fixed time-like vector.
If, following \cite{Myers:2003fd} and \cite{Gubitosi:2009eu}, we write the coupling constant between the EM field and the vector through a dimensionless parameter $\xi$ and the Planck mass scale $M_{P}$, then the polarization rotation angle is again related in a linear way to the distance covered by photons when redshift effects are disregarded:
\begin{equation}
\alpha(E)=\frac{\xi}{M_{P}}E^{2}\Delta\ell.
\end{equation}
Taking redshift into account this becomes:
\begin{equation}
\alpha(E_{0},z_\star)=\frac{\xi}{M_{P}}E_{0}^{2}\int_{0}^{z_{\star}}(1+z)H(z)^{-1}dz\,.
\end{equation} 
where $z_\star$ and $E_0$ are defined as before.
Using this last formula for our analysis, the best-fit estimate for $\xi$ is:
\begin{equation}
\xi=(1.2\pm14.1)\cdot 10^{-11}\,,
\end{equation}
at 68\% C.L.  including all data points except GRB.
Differently from what happened in the linear energy dependence, now it is actually the highest energy source
(Crab Nebula) that gives the most important contribution, weighing the energy of the source more that its distance.
If we include also the constraint from GRB we obtain at 68\% C.L.:
\begin{equation}
\xi=(0.0\pm8.6)\cdot 10^{-17}\,.
\end{equation}
As expected, the GRB provides the dominant contribution and our result is indeed compatible with the upper limit presented in \cite{Gotz:2013dwa}.
Again, in both cases the result improves the constraint, $\xi=(-0.22\pm 0.22)$ at  68\% C.L., based only 
on CMB dataset \cite{Gubitosi:2012rg} by several orders of magnitude.

\section{Conclusions}
In the present work we have constrained several non-standard electromagnetic theories which predict a rotation of the photon polarization plane (cosmological birefringence). Each of the models considered predicts a different  energy and distance dependence of the effect, which must be taken into account when combining results from observations of different kinds of sources. We have done so, and combined for the first time the constraints on the rotation angle set by cosmological (CMB) and astrophysical (UV distant radio galaxies, radio sources, Crab Nebula, GRBs) observations. Besides updating current constraints on the models considered, our analysis also provides a useful guide for future 
polarization measurements at different frequencies 
aimed at investigating specific energy- and distance-dependent birefringence effects.

\section*{Acknowledgements}
We wish to thank Paddy Leahy for useful discussions on the polarization conventions.
GG was supported in part by the John Templeton Foundation. GG wishes to thank SISSA for hospitality during the development of this work;
MG thanks INAF-IASF Bologna and Vatican Observatory for hospitality.

%\bibliographystyle{apsrev4-1}
%\bibliography{bibliography} 

\end{document}